\documentclass[10pt, letter]{IEEEtran}

\usepackage{amssymb}
\usepackage{epsfig,latexsym}
\usepackage{graphicx}
\usepackage{color}
\usepackage{graphicx}
\usepackage{subfigure}
\usepackage{algorithm}
\usepackage{algorithmic}
\usepackage{stmaryrd}
\usepackage{mathrsfs}
\usepackage{amsfonts}
\usepackage{amsmath}
\usepackage{cases}
\usepackage{txfonts}
\usepackage{amssymb}
\usepackage{epsfig}
\usepackage{bm}
\usepackage{caption}
\usepackage{cite}
\usepackage{epstopdf}

\begin{document}

\title{Constrained Coding and Deep Learning Aided Threshold Detection for Resistive Memories}

\author{Xingwei Zhong, Kui Cai, Guanghui Song, Weijie Wang, and Yao Zhu \vspace{-0.4cm}}

\maketitle

\begin{abstract}
Resistive random access memory (ReRAM) is a promising emerging
non-volatile memory (NVM) technology that shows high potential for
both data storage and computing. However, its crossbar array
architecture leads to the sneak path problem, which may severely
degrade the reliability of data stored in the ReRAM cell.  Due to
the complication of memory physics and unique features of the
sneak path induced interference (SPI), it is difficult to derive
an accurate channel model for it. The deep learning (DL)-based
detection scheme \cite{zhong2020sneakdl} can better mitigate the
SPI, at the cost of additional power consumption and read latency.
In this letter, we first propose a novel CC scheme which can not
only reduce the SPI in the memory array, but also effectively
differentiate the memory arrays into two categories of
sneak-path-free and sneak-path-affected arrays. For the
sneak-path-free arrays, we can use a simple middle-point threshold
detector to detect the low and high resistance cells of ReRAM. For
the sneak-path-affected arrays, a DL detector is first trained
off-line (prior to the data detection of ReRAM). To avoid the
additional power consumption and latency introduced by the DL
detector, we further propose a DL-based threshold detector, whose
detection threshold can be derived based on the outputs of the DL
detector. It is then utilized for the online data detection of all
the identified sneak-path-affected arrays. Simulation results
demonstrate that the above CC and DL aided threshold detection
scheme can effectively mitigate the SPI of the ReRAM array and
achieve better error rate performance than the prior art detection
schemes, without the prior knowledge of the channel.

\end{abstract}

\section{Introduction}

Owing to its superior features of simple structure, fast
read/write speed, low power consumption, and high scalability, the
resistive random access memory (ReRAM) is widely considered a
promising non-volatile memory (NVM) candidate to replace the flash
memory \cite{yuj}. By adjusting the write voltage to the ReRAM
cell, its resistance can be programmed to two resistance states: a
Low-Resistance State (LRS) representing an input information bit
of `1', and a High-Resistance State (HRS) denoting an input
information bit of `0'. A key feature of ReRAM is that the memory
cell can be both written and read over a simple crossbar array,
which can achieve a significant density gain. However, it also
causes a serious problem known as the sneak path \cite{cassutoj}.
Sneak paths are undesired paths in parallel to the desired read
path, which cause interference to the cell being read and hence
the read errors. A distinct feature of the sneak path induced
interference (SPI) is that it is data-dependent. That is, the
occurrence of the sneak path depends on the pattern of the input
data \cite{cassutoj,chenj,songj}.

A common method to mitigate the SPI is to place a cell selector in
series to each array cell which only allows the current to flow in
one direction. However, due to imperfections in the memory
fabrication and maintenance processes, the cell selectors are also
prone to failures \cite{chenj,songj}, and hence the SPI may still
occur. In \cite{cassutoj}, the SPI is modelled as parallel
resistances of a probabilistic sneak path model, and data
detection schemes are proposed by treating the SPI as an
independent and identically distributed ({\it i.i.d.}) noise. The
inter-cell correlation introduced by the SPI is utilized by a
later work of \cite{chenj}, where several joint-cell data
detection schemes are presented by introducing the diagonal-0
cells. In particular, the inserted diagonal-0 cells can provide
side information for determining the detection thresholds for the
so-called Double Threshold Scheme and Triple Threshold Scheme. In
\cite{songj}, an elementary signal estimator (ESE) is proposed,
which can estimate the ratio of the HRS cells that are affected by
the SPI, and further generate both the hard detected bits and the
soft information of the ReRAM channel.

All of the above prior-art detection schemes
\cite{cassutoj,chenj,songj} are model-specific. That is, the
detectors assume that the full knowledge of the ReRAM channel is
known and matched with the channel models they adopted. However,
in practice, due to the complicated nature of the SPI as well as
the complication of memory physics, it is difficult to derive an
accurate channel model of the SPI. All the sneak path models
adopted by the prior-art detectors have simplified the SPI
significantly, and hence may not match exactly with that in the
practical ReRAM array. To overcome the limitation of these
detectors, a deep learning (DL)-based detection scheme is proposed
in \cite{zhong2020sneakdl}, which is data-driven, and hence does
not rely on the prior knowledge of the ReRAM channel. A data
preprocessing scheme \cite{zhong2020sneakdl} is also proposed for
the DL detector, to differentiate memory arrays into two
categories of without and with SPI. It is based on the observation
that not all the memory arrays are corrupted by the SPI. Hence if
using a mixed set of data collected from various arrays as the
training set of the deep neural network (DNN), the obtained NN
parameters will lead to inferior detection performance for both
types of arrays without and with the SPI. A manually-tuned
resistance threshold is required for the data pre-processor to
differentiate the memory arrays \cite{zhong2020sneakdl}, where the
tuning process is time-consuming and may not accurately
differentiate the memory arrays. Furthermore, the DL detector is
adopted for both type of memory arrays in \cite{zhong2020sneakdl},
which results in a significant increase of the power consumption
and read latency.

In this letter, we propose a novel constrained coding (CC) and DL
aided threshold detection scheme for ReRAM channel with SPI. In
particular, in Section II, we introduce the ReRAM channel model
that is used to generate data for training and testing the DL
detector. In Section III, we propose a novel CC scheme which can
not only reduce the SPI in the memory array, but also effectively
differentiate memory arrays into two categories of sneak-path-free
and sneak-path-affected arrays. For the sneak-path-free arrays, we
can use a simple middle-point threshold detector to differentiate
the low and high resistance cells. For the sneak-path-affected
arrays, a DL detector is first trained off-line. In order to avoid
the additional power consumption and latency introduced by the DL
detector, we further propose a DL-based threshold detector in
Section IV, whose detection threshold can be derived based on the
outputs of DL detector. It is then utilized for the online data
detection of all the identified sneak-path-affected arrays. Thus,
we avoid using the DL detector for both memory arrays without and
with SPI. Simulation results illustrated in Section V demonstrate
that our proposed CC and DL aided threshold detection scheme can
effectively mitigate the SPI and achieve better bit error rate
(BER) performance than the prior art detection schemes, without
the prior knowledge of the channel. The letter concludes in
Section VI.

\section{The ReRAM Channel Model}
We consider an $N \times N$ ReRAM crossbar array which contains
$N^2$ resistive cells. The cell $(i,j)$ that lies at the
intersection of row $i$ and column $j$ stores binary bit
$A_{i,j}$, where $i,j \in \{1,\cdots, N\}$. By applying an
external voltage pulse across the memristor cell, it enables a
transition of the device from a LRS to a HRS and vice versa.
During the read operation, certain voltage is applied to a target
cell $(i,j)$ to measure its resistance. In parallel to the desired
measurement path, the sneak path is the alternative current path
that originates from and returns back to the target cell $(i,j)$
while traversing other memory cells with LRS through alternating
horizontal steps and vertical steps \cite{songj}. As a result, the
measured resistance of the target cell will be decreased. Hence
this detrimental effect only occurs when a logical `0' bit (cell
with HRS) is read.

In the actual crossbar arrays, the most common way to mitigate the
SPI is by adding the cell selectors. A selector is an electrical
device which only allows the current to flow in one direction. As
the sneak path inherently causes the reverse current in at least
one of the cells locating along the parallel path, introducing a
selector in series to each memory cell can completely eliminate
the SPI in the entire memory array \cite{cassutoj}. However, due
to the imperfections in the production or maintenance of the
memory, the cell selectors may also fail, leading to the
reoccurrence of the SPI. Following \cite{cassutoj,chenj,songj}, we
only consider the most dominant sneak path of length 3. The
failure probability of the selector is {\it i.i.d} with
probability $p_f$ \cite{cassutoj,chenj,songj}. Generally, target
cell $(i,j)$ is affected by the SPI if the following three
conditions are all satisfied:

1) The target cell $(i, j)$ is in HRS, {\it i.e.}, $A_{i,j} = 1$.

2) We can find at least one combination of $u,v \in[1,\cdots,N], u
\neq i, v \neq j$ that induces a sneak path, denoted by $A_{i,v} =
A_{u,v} = A_{u,j} = 1$.

3) The cell selector fails at the diagonal cell $(u,v)$.

According to \cite{cassutoj,chenj,songj}, the sneak path affected
ReRAM channel can be described as:
\begin{equation} \label{sneakpathmodel}
\begin{aligned}
&r_{{i,j}}=R_{{i,j}}+\eta_{{i,j}} \\
&with~R_{{i,j}}=
   \begin{cases}
   R_1~~~~~~~~~~~~~~~if~A_{i,j}=1,\\
   \left( \dfrac{1}{R_0}+\dfrac{e_{{i,j}}}{R_{sp}}\right) ^{-1}~~if~A_{i,j}=0,
   \end{cases}
\end{aligned}
\end{equation}
where $r_{{i,j}}$ is the measured resistance value of an ReRAM
cell, and $e_{{i,j}}$ is a boolean random variable where
$e_{{i,j}}$ = `1' (or `0') indicates whether or not a HRS cell at
position $(i,j)$ is affected by the SPI. Here, $R_{sp}$ denotes
the parasitic resistance caused by the sneak path, and $R_1$ and
$R_0$ represent the nominal values of the LRS and HRS,
respectively. We also include a Gaussian distributed noise
$\eta_{i,j}$ with zero mean and variance $\sigma^{2}$, to model
the combined effect of various other noises within the memory
system.

\textbf{Remark 1:} Note that the channel model given by
(\ref{sneakpathmodel}) is only used to generate the necessary data
({\it i.e.}, the resistance values $r_{i,j}$ and the corresponding
labels $A_{i,j}$) for training and testing the DL detector. Our
subsequently proposed CC and DL aided threshold detection scheme
does not need to have the prior knowledge of the channel.

\section{Constrained Coding-Aided Deep Learning Detection}
Due to the complicated nature of the SPI ({\it e.g.} both the
number and structure of the sneak paths are data-dependent) as
well as the complication of memory physics, it is difficult to
derive an accurate channel model of the SPI. The ReRAM channel
model defined by (\ref{sneakpathmodel}) only considers a single
sneak path in the memory array, while in practice there may be
multiple sneak paths. The DL-based detection scheme proposed in
\cite{zhong2020sneakdl} can overcome this limitation, as it is
data-driven and does not rely on the prior knowledge of the ReRAM
channel. During the training of the DL detector for ReRAM, it is
observed that not all the memory arrays are corrupted by the SPI
\cite{zhong2020sneakdl}. If all the data collected from various
arrays is utilized as a whole training set of the DNN, the
obtained NN parameters will lead to inferior detection performance
for both memory arrays without and with the SPI. Hence a data
preprocessing scheme is presented in \cite{zhong2020sneakdl},
which adopts a manually-tuned resistance threshold to
differentiate the memory arrays into two types: the
sneak-path-free arrays and sneak-path-affected arrays. However,
tuning the resistance threshold manually is time-consuming, and it
may not accurately differentiate the memory arrays. In this work,
we propose a novel CC scheme which can not only differentiate the
memory arrays accurately without involving the manually tuning
process, but also minimize the SPI occurred in the crossbar array.

\begin{figure}[t]
\centering
\includegraphics[width=3.5in, height=1.8in]{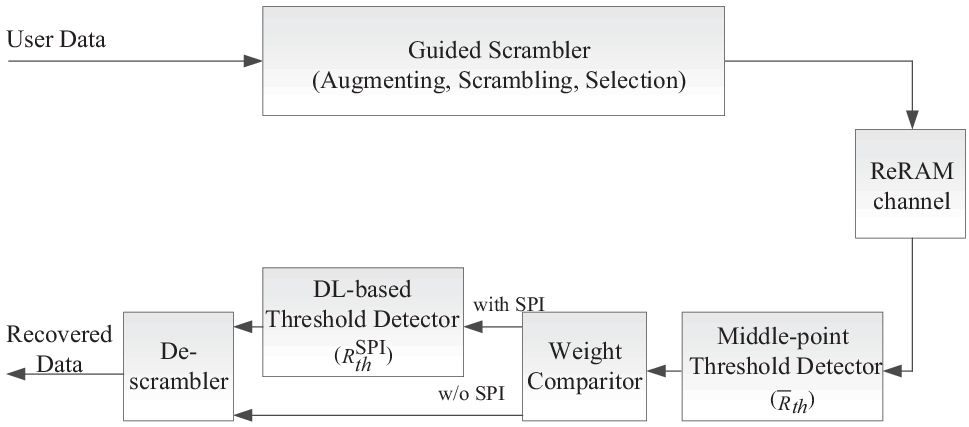}
\caption{Block diagram of CC and DL aided threshold detection for the ReRAM channel.}
\label{gm_whole_structure}
\end{figure}

As shown by Fig. \ref{gm_whole_structure}, the proposed CC method
for ReRAM is based on the guided scrambling (GS) technique
\cite{fair1991guided,immink1997performance}. GS is a simple and
efficient coding technique that is widely used in designing the
dc-free constrained codes, the spectrum shaping codes, the run
length limited (RLL) codes, and their combinations [Immink 2012
paper] [My CE 2017 paper]. In this work, we apply the GS technique
to design constrained codes to minimize the SPI. The encoding
consists of the following three steps.

\textbf{Augmenting}: First to reduce the computational complexity,
we divide the input data array $A$ of size $N \times N$ into
several sub-arrays of size $M \times M$, with $M<N$. By further
augmenting the input user data bits $\pmb{a} =\lbrace
A_{1,1},\cdots,A_{M,t} \rbrace$ with $l$ redundant bits, $t=\mod
(M^2-l,M)$, an intermediate set $\pmb{I}$ can be generated.  The
intermediate set $\pmb{I}= \lbrace I_{1},\cdots,I_{2^{l}} \rbrace$
with $2^{l}$ candidate codewords can be written as:

\begin{equation} \label{candidate_codewords} \nonumber
\begin{aligned}
&I_{1}=\left[
\begin{matrix}
A_{1,1} & \cdots & \cdots & A_{1,M}\\
\vdots &\multicolumn{2}{c}{\ddots} & \vdots\\
A_{M,1} & \cdots & A_{M,t} & 0 \cdots 0\\
\end{matrix}
\right]_{M\times M}, \\
&I_{2}=\left[
\begin{matrix}
A_{1,1} & \cdots & \cdots & A_{1,M}\\
\vdots &\multicolumn{2}{c}{\ddots} & \vdots\\
A_{M,1} & \cdots & A_{M,t} & 0 \cdots 1\\
\end{matrix}
\right]_{M\times M}, \\
& \qquad \qquad \qquad \ \  \cdots  \\
&I_{2^{l}}=\left[
\begin{matrix}
A_{1,1} & \cdots & \cdots & A_{1,M}\\
\vdots &\multicolumn{2}{c}{\ddots} & \vdots\\
A_{M,1} & \cdots & A_{M,t} & 1\cdots 1\\
\end{matrix}
\right]_{M\times M}.
\end{aligned}
\end{equation}

\textbf{Scrambling}: Next, a feedback register scrambler
\cite{fair1991guided} scrambles all the matrices in $\pmb{I}$ to
obtain the selection set $\pmb{S} =\lbrace S_{1},\cdots,S_{2^{l}}
\rbrace$, given by
\begin{equation} \label{scrambler}
S_{k} = I_{k}\oplus \sum_{p\in C}^{}S_{k-p},
\end{equation}
where $\oplus$ is the XOR operation, and $C$ is the set of
non-zero coefficients of the scrambler polynomial
$g(x)=x^r+\sum_{i=1}^{r}c_ix_{r-i}$, $c_i \in \{0,1\}$
\cite{fair1991guided}.

\textbf{Selection}: In GS, the criterion to select the `best' word
from the selection set is key to achieve the desired code
constraint. In this work,
we propose a {\it minimum number of sneak paths} (MNSP) criterion
to design the minimum-SPI constrained code for the ReRAM channel.
That is, among the matrices in the selection set $\pmb{S}$,
we choose the matrix $S_{c}$ that will result in the minimum
number of possible sneak paths in the corresponding memory
array. We define a possible sneak path when the current
memory cell is in HRS, and three other cells at the
corners of a rectangular path formed by these four cells
are all in LRS. Obviously, such a path will satisfy the
first two conditions for the sneak path to occur as we
described in Section II, and hence a sneak path may occur
depending whether the third condition is satisfied. Therefore,
the selected matrix $S_c$ is considered as a minimum-SPI
constrained code.\\
\indent After that, the selected matrix $S_c$ that will
lead to the minimum possible number of sneak paths is
written into the memory array. Meanwhile, the weight of
the selected matrix $S_c$ ({\it i.e.}, sum of all the binary
entries of $S_c$) is stored by the system also. For
decoding, the de-scrambler will recover the input user
word by imposing an inverse operation to the data of
output memory array, given by $I_{k} = S_{k}\oplus \sum_{p\in C}^{}I_{k-p}$.
The rate of the designed constrained code is thereby $R_{c}=\frac{M^2-l}{M^2}$.

At the receiver side, as shown by Fig. 1, the signals read back
from the ReRAM channel are first sent to a conventional
middle-point threshold detector. That is, the resistance threshold
of the detector is taken as ${\bar R}_{th}=(R_0+R_1)/2$. The
weight of the detected bits of the memory array is then compared
with that of the transmitted array stored by the system through a
weight comparator. If they are the same, the memory array is
considered to be SPI-free, and the detected array bits are sent to
the de-scrambler directly. Otherwise, the memory array is regarded
as SPI-affected, and a DL detector will be used for data
detection. Moreover, it will be trained by the data associated
with the identified SPI-affected arrays only. Therefore, the above
described classification of the memory arrays can be considered as
array-weight-based data preprocessing for the DL detector. The
architecture of the DNN \cite{zhong2020sneakdl} consists of 4
layers: an input layer with $N^2$ nodes, two hidden layers with $4
N^2$ and $2 N^2$ nodes, respectively, and an output layer with
$N^2$ nodes. The resistance values $\pmb{r}$ read back from the
memory array are the inputs to the NN, while the NN outputs soft
estimates $\pmb{\tilde{A}}$ of the label $\pmb{A}$, based on which
the hard decisions $\pmb{\hat{A}}$ can be obtained. It is then
sent to the de-scrambler to recover the original user data.

\textbf{Remark 2:} The above proposed method for designing the
minimum-SPI constrained codes can be used separately without
associated with the DL detector. The constrained codes proposed in
the literature for ReRAM all aim at reducing the number of LRS
cells ({\it i.e.} the number of `1's in the codeword) to suppress
the SPI \cite{cassutoj}. In this work, we propose the (MNSP)
criterion for GS to minimize the SPI directly for each constrained
codeword written into the memory array. As will be shown by the
simulation results in Section IV, our proposed constrained codes
can reduce the SPI more effectively than enforcing the minimum
weight constraint.

\textbf{Remark 3:} Simulation results in Section V show that the
above CC-aided method can identify the memory arrays without and
with SPI accurately. Hence as shown by Fig. 1, to detect the
sneak-path-free arrays, unlike the work of
\cite{zhong2020sneakdl}, we can just adopt a simple middle-point
threshold detector instead of a DL detector (since for ReRAM, the
values of $R_1$ and $R_0$ are far from each other \cite{yuj,
cassutoj,chenj,songj}). For the sneak-path-affected arrays, a
better classifications of the memory array will facilitate a
better training of the NN parameters and hence a better BER
performance of the DL detector.

\section{Deep Learning-Based Threshold Detection and Bit Error Rate Bound}
As described in the previous section, a DL detector is adopted for
the sneak-path-affected arrays. Although the DL detector for the
ReRAM channel can achieve superior performance, it needs to be
activated for each input data sequence. This will lead to a
significant increase of the power consumption and the read
latency. Therefore, we develop a DL-based threshold detector for
sneak-path-affected arrays, whose detection threshold can be
derived based on the outputs of DL detector. The details are as
follows.

First, based on the output of the NN detector ${\tilde {A}_{i,j}}$
which give the soft estimate of the binary bit ${{A_{i,j}}}$
stored in cell $(i,j)$, the hard estimation ${\bar {A}_{i,j}}$ of
${{A_{i,j}}}$ can be obtained using the following hard decision
rule: if ${\bar {A}_{i,j}}>0.5$, ${\bar {A}_{i,j}}=1$; otherwise
${\bar {A}_{i,j}}=0$. On the other hand, with an assumed detection
threshold $R_{th}$, we can get the hard estimation ${\hat
{A}_{i,j}}$ for a given memory cell readback signal $r_{_{i,j}}$.
Hence we can obtain an adjusted detection threshold
${R}_{th}^\text{SPI}$, by searching for the threshold $R_{th}$
that minimizes the Hamming distance between ${\bar {A}_{i,j}}$ and
${\hat {A}_{i,j}}$, over a large number $T$ of memory arrays. We
thus have
\begin{equation} \label{R_pred}
R^{\text{SPI}}_{th}=\text{arg}\min_{R_{th}}\sum_{i=1}^{T}
d(\pmb{\bar{A}}^{i}_{{R}_{th}}, \pmb{\hat{A}}^{i}).
\end{equation}

Note that in practice, the DL detector for the sneak-path-affected
arrays will first be trained off-line, prior to the data detection
of ReRAM arrays. The above derivation of the adjusted detection
threshold ${R}_{th}^\text{SPI}$ based on the outputs of the DL
detector will also be carried out off-line. The obtained DL-based
threshold detector is then utilized for the online data detection
of all the sneak-path-affected arrays identified by CC, as shown
by Fig. 1. In this way, we avoid the use of the DNN detector for
both memory arrays without and with SPI. This will lead to a
significant reduction of the power consumption and read latency.

In addition, to provide a reference for evaluating the performance
of the various detection schemes, we also derive a lower bound of
BER for the ReRAM channel with SPI. In particular, for an $N
\times N$ ReRAM crossbar array with the input distribution to be
{\it i.i.d.} Bernoulli $(q)$, {\it i.e.},
$\textrm{Pr}(A_{i,j}=1)=q$ and $\textrm{Pr}(A_{i,j}=0)=1-q$, the
probability of a memory cell not being affected by the SPI is
given by:
\begin{equation} \label{non-sneak path}
\begin{aligned}
&P_{nonsp}=\sum_{u=0}^{N-1}\sum_{v=0}^{N-1}\binom{N-1}{u}
\binom{N-1}{v} q^{u+v}(1-q)^{N-1-u+N-1-v} \\
&\  \  \ \ \  \  \ \ \  \ (1-p_f \times q)^{uv}.
\end{aligned}
\end{equation}
Then an lower bound of the overall BER can be written as:
\begin{equation} \label{non-selector failure bound}
\begin{aligned}
P_{BER}=P_{nonsp} Q(\dfrac{R_0-R_1}{2\sigma})+ (1-P_{nonsp})
Q(\dfrac{R_{0\_sp}-R_1}{2\sigma}),
\end{aligned}
\end{equation}
where $Q(x) = \dfrac{1}{2\pi}\int_{x}^{ \infty }exp(-\dfrac{u^{2}}{2})du $ is the Q-function.

\begin{figure}[b]
\centering
\includegraphics[width=3.2in, height=2.4in]{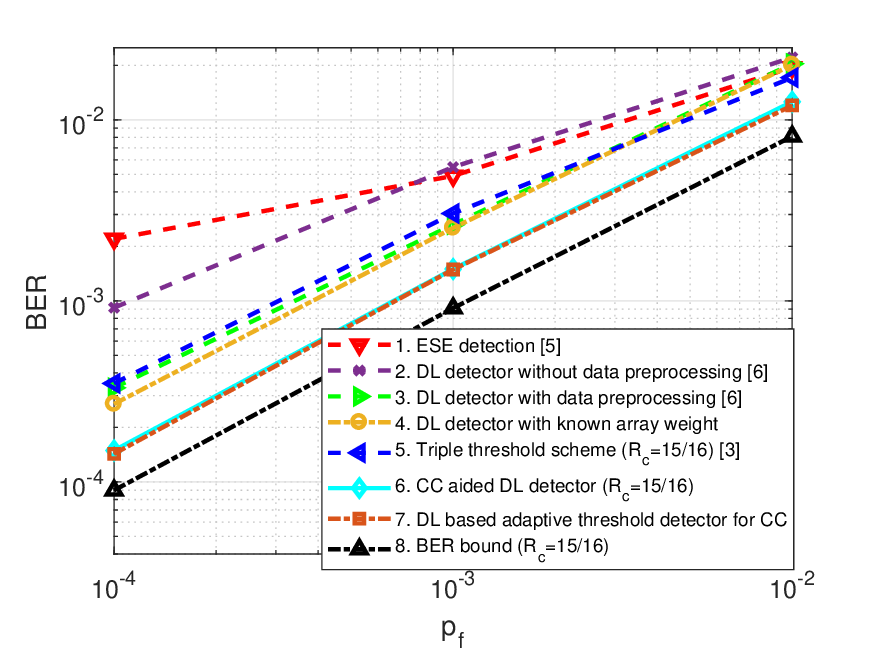}
\caption{BER comparison of different detectors over different $p_f$, with $\sigma=30$.}
\label{precoding_with_different_pf_sigma_30}
\end{figure}

\begin{figure}[t]
\centering
\includegraphics[width=3.0in, height=2.2in]{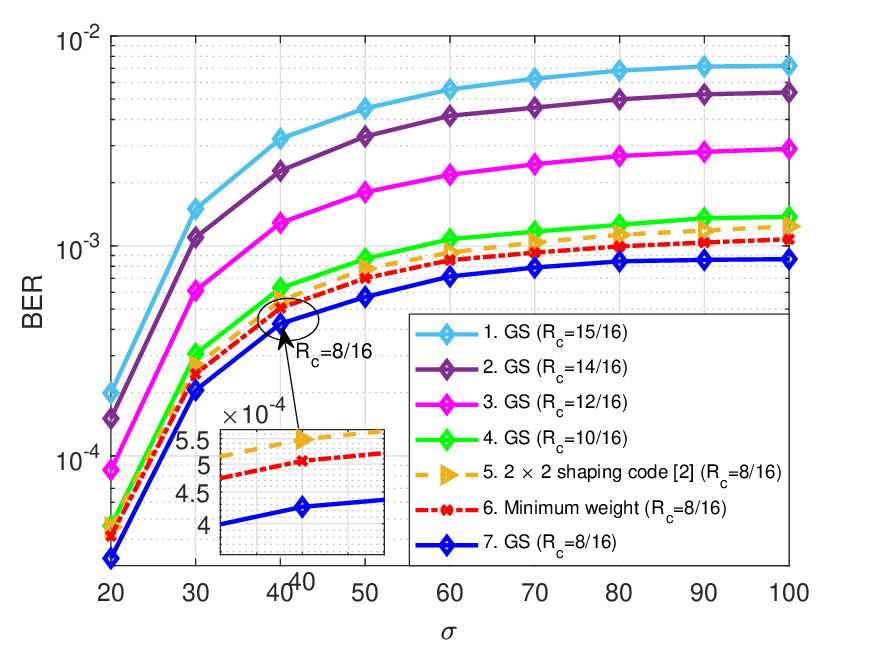}
\caption{BER comparison of CC and DL aided threshold detector
with different CC schemes, with $p_f=10^{-3}$.}
\label{precoding_case_8_16}
\end{figure}

\section{Performance Evaluations}
Computer simulations are carried out to evaluate the performance
of different detection schemes. In the simulations, we follow the
literature \cite{cassutoj,chenj,songj} and assume that
$R_1=100\Omega$, $R_0=1000\Omega$, and $R_{sp}=250\Omega$. The
array size is set to be $N\times N=16 \times 16$.  For the
DL-based detectors, we adopt the same network settings as
\cite{zhong2020sneakdl}, with training samples $1\times 10^6N$,
testing samples $5\times 10^6N$, mini-batch size $4N^2$, the Adam
optimizer, the rectified linear unit (ReLU) and the sigmoid
activation function, and the binary cross-entropy (BCE) loss
function.

We first take a selector failure rate of $p_f=10^{-3}$ by
following the literature \cite{cassutoj,chenj,songj}, and evaluate
the BERs of different detectors over different noise magnitude
$\sigma$. The results are illustrated by Fig.
\ref{precoding_different_sigma}. Since in \cite{chenj}, the
insertion of the diagonal-0 cells leads to a code rate of
$R_c=(N-1)/N=60/64=15/16$, the rate of our CC scheme is also kept
the same. Correspondingly, we set $l=4$, $M=8$, and
$g(x)=1+x+x^{4}$. All the rest detectors will not incur coding
redundancy.

From Fig. \ref{precoding_different_sigma}, we observe that at high
noise regions of $\sigma>40$, the ESE detector \cite{songj} (Curve
1) performs the worst among all the detectors. The DL detector
developed by \cite{zhong2020sneakdl} without data preprocessing
(Curve 2) outperforms the ESE detector. The DL detector with the
resistance-threshold-based data preprocessing
\cite{zhong2020sneakdl} (Curve 3) achieves better performance than
the case without data-preprocessing. However, it BERs are higher
than the Triple Threshold Scheme \cite{chenj} (Curve 5). At noise
regions of $\sigma<40$, it outperforms the other three detectors.
Note that for this case, the resistance threshold that is used to
differentiate the memory arrays has to be tuned manually for each
of the noise magnitude.

It can be further seen from Fig. \ref{precoding_different_sigma}
that our proposed CC-aided DL detector (Curve 6) performs
significant better than all the above detectors, and over a wide
range of noise levels. The performance gain is attributed to both
the array-weight-based data preprocessing and the SPI reduction.
To illustrate the performance gain in more details, we include
Curve 4 which shows the performance of the DL detector with the
array-weight-based data preprocessing only and without CC.
Therefore, the gap between Curve 2 and Curve 4 illustrates the
performance gain due to the array-weight-based data preprocessing,
while the gap between Curves 4 and 6 demonstrates the BER gain due
to the suppression of SPI. Furthermore, it is also observed that
the proposed CC and DL aided threshold detector (Curve 7) achieves
very similar performance with the CC-aided DL detector (Curve 6),
thus leading to a significant reduction of the power consumption
and read latency. Meanwhile, it still has a noticeable gap with
the BER bound (Curve 8). This is due to the fact that the SPI is
data-dependent, and hence the level of SPI varies among different
memory arrays depending on the different input data arrays, which
is unknown to the channel detector. Hence the NN parameters of the
DL detector trained by a large number of memory arrays (each with
different levels of SPI) can only lead to an averaged BER
performance of the DL detector.

Next, we evaluate the performance of the above detectors over
different selector failure rate $p_f$, while keeping a medium
noise level of $\sigma=30$. Fig.
\ref{precoding_with_different_pf_sigma_30} shows that our proposed
CC and DL aided threshold detector achieves similar performance
with the CC-aided DL detector for all the different selector
failure rates, with much less power consumption and read latency.
Both detectors perform significantly better than the prior art
detectors. In addition, the comparison between Curve 3 and Curve 4
shows that our CC-aided DL detector can differentiate the array
types more accurately (by using the weight of the received memory
data array) than the DL detector of \cite{zhong2020sneakdl} which
adopts the manually tuned resistance threshold to classify the
memory arrays. The difference is larger at low selector failure
rate regions. Moreover, we also verified that Curve 4 overlaps
with the BERs of the ideal case that the DL detector is trained
with the exact sneak-path-affected arrays, which indicates that
our CC-aided DL detector can differentiate the array types
accurately for different selector failure rates.

Lastly, we focus on the CC and DL aided threshold detector and
evaluate its performance with different CC schemes, with
$p_f=10^{-3}$. The simulation results are illustrated by Fig.
\ref{precoding_case_8_16}. For our proposed CC scheme with GS and
the MNSP criterion, we change the code rate by adjusting the
number of redundant bits $l$, the sub-arrays number $M$, and the
scrambler polynomial $g(x)$. For $R_c=56/64=14/16$, we have $l=8$,
$M=8$, and $g(x)=$. For the cases of $R_c=12/16, 10/16, 8/16$, we
have $M=4$, $l=4,6,8$ and $g(x)=$, respectively. We observe from
Fig. \ref{precoding_case_8_16} that by introducing more redundant
bits and reducing the codes rates, the BER performance is improved
progressively. Moreover, for the case with $R_c=8/16$, we are able
to compare its performance with a prior art $2 \times 2$ shaping
code \cite{cassutoj}, since their code rates are the same. We also
include the BERs of the CC scheme with GS and the same code rate,
but with the minimum weight (of the data array) selection
criterion. Fig. \ref{precoding_case_8_16} clear shows that our
proposed CC scheme outperform both the $2 \times 2$ shaping code
and the code designed using the minimum weight selection
criterion.

\section{Conclusions}

To combat the SPI that severely affects the performance of the
ReRAM channel, we have proposed a novel CC and DL aided threshold
detection scheme without the prior knowledge of the channel. In
particular, we have first proposed a CC method which can not only
choose the best constrained codeword to reduce the SPI, but also
differentiate two types of arrays effectively for the data
preprocessing of the DL detector. To avoid the additional power
consumption introduced and latency caused by the DL detector, we
have further proposed a DL-based threshold detection scheme, whose
detection threshold can be derived based on the outputs of the DL
detector. We have also derived  a lower bound of BER for the ReRAM
channel with SPI to benchmark the performance of the proposed
detectors. Various simulation results demonstrate that the
proposed CC and DL aided threshold detection scheme can
effectively mitigate the SPI without the prior knowledge of the
channel, and hence it shows great potential for high-density ReRAM
crossbar arrays.


\begin{thebibliography}{1}

\bibitem{yuj}S. Yu, and P. Y. Chen, ``Emerging memory technologies:
recent trends and prospects," \emph{IEEE Solid-State Circuits
Mag.}, vol.~8, no.~2, pp.~43--56, 2016.

\bibitem{cassutoj}Y. Ben-Hur, and Y. Cassuto, ``Detection and
coding schemes for sneak-path interference in resistive memory
arrays," \emph{IEEE Trans. Commun.}, vol.~67, no.~6,
pp.~3821--3833, 2019.

\bibitem{chenj}Z. Chen, C. Schoeny and L. Dolecek, ``Pilot
assisted adaptive thresholding for sneak-path mitigation in
resistive memories with failed selection devices," \emph{IEEE
Trans. Commun.}, vol.~68, no.~1, pp.~66--81, 2020.

\bibitem{songj}G. Song, et al., ``Performance limit and
coding schemes for resistive random-access memory channels,"
\emph{IEEE Trans. Commun.}, vol.~69, no.~4, pp.~2093--2106, 2021.

\bibitem{zhong2020sneakdl}X. Zhong, K. Cai, G. Song and N.
Raghavan, ``Deep learning based detection for mitigating sneak
path interference in resistive memory arrays," in \emph{Proc. IEEE
International Conference on Consumer Electronics - Asia
(ICCE-Asia), Seoul, Korea (South)}, Nov. 2020, pp.~ 1--4.

\bibitem{fair1991guided}I.J. Fair, W.D. Grover, W.A. Krzymien,
and R. I. MacDonald, ``Guided scrambling: a new line coding
technique for high bit rate fiber optic transmission systems,"
\emph{IEEE Trans. Commun.}, vol.~39, no.~2, pp.~289--297, 1991.

\bibitem{immink1997performance}K. A. S. Immink and L. Patrovics,
``Performance assessment of DC-free multimode codes," \emph{IEEE
Trans. Commun.}, vol.~45, no.~3, pp.~293--299, 1997.

\end{thebibliography}
\end{document}